\documentclass[sigconf,nonacm]{acmart}
\AtBeginDocument{%
  }

\usepackage{amsmath,amsfonts}
\usepackage{algorithmic}
\usepackage{subcaption}
\usepackage{graphicx}
\usepackage{textcomp}
\usepackage{xcolor}
\usepackage{wasysym}
\usepackage{listings}
\usepackage{float}
\usepackage{url}
\usepackage{xspace}
\usepackage[table]{xcolor}

\begin{document}

\title{Hojabr: Towards a Theory of Everything for AI and Data Analytics}

\author{Amir Shaikhha}
\email{amir.shaikhha@ed.ac.uk}
\affiliation{%
  \institution{University of Edinburgh}
  \city{Edinburgh}
  \country{United Kingdom}
}


\newcommand{\smartpara}[1]{\noindent \textbf{#1.}}

\newcommand{\RA}[1]{{\color[HTML]{0088AA}{#1}}}
\newcommand{\RB}[1]{{\color[HTML]{AA4499}{#1}}}
\newcommand{\RC}[1]{{\color[HTML]{D55E00}{#1}}}
\newcommand{\RAll}[1]{{\color[HTML]{117733}{#1}}}

\colorlet{myblue}{blue!50!black}
\colorlet{mygreen}{green!50!black}

\newcommand{\lang}{Hojabr\xspace}

\newcommand{\amirsh}[1]{{\color{red}Amir: #1}}
\newcommand{\todo}[1]{{\color{red}TODO: #1}}

\newcommand{\changed}[1]{{\color{blue}#1}}

\newcommand{\codekwstyle}{\ttfamily\bfseries\color{myblue}}
\newcommand{\codemcstyle}{\ttfamily\bfseries\color{mygreen}}

\newcommand{\code}[1]{{\texttt{#1}}}
\newcommand{\codekw}[1]{{\codekwstyle\texttt{#1}}}
\newcommand{\codemc}[1]{{\codemcstyle\texttt{#1}}}
\newcommand{\codetiny}[1]{{\footnotesize\ttfamily\texttt{#1}}}
\newcommand{\codekwpy}[1]{{\footnotesize\color{blue}#1}}

\newcommand{\ruleaction}{$\lhd$}

\definecolor{colg}{rgb}{0.1,0.7,0.1}
\definecolor{colr}{rgb}{0.7,0.1,0.1}
\definecolor{colb}{rgb}{0.1,0.1,0.7}
\definecolor{colbb}{rgb}{0.1,0.1,0.6}
\newcommand{\supfull}{{\color{colg}{\CIRCLE}}}
\newcommand{\suphalf}{{\color{colg}{\LEFTcircle}}}
\newcommand{\supnone}{\Circle}


\lstdefinelanguage{hojabr}{
  keywords={if,or,true,false,not,in,%
    avg,sum,min,max,median,%
    limit,order,top,%
    type,card,deg,unique,and,%
    pkey,fdep,%
    sin,cos,relu,softmax,%
    match,case,while
    },%
  morekeywords={UID, round},%
  emph={CEI,EEI},
  sensitive,%
  morecomment=[l]//,%
  morecomment=[s]{\{}{\}},%
  morestring=[b]",%
  showstringspaces=false,%
  breaklines=true,%
  mathescape=true,%
  showspaces=false,
  showtabs=false,
  showstringspaces=false,
  breakatwhitespace=true,
  xleftmargin=1em,
  aboveskip=1pt,
  belowskip=1pt,
  lineskip=-0.2pt,
  basicstyle=\ttfamily,
  keywordstyle=\codekwstyle{},%
  emphstyle=\codemcstyle{},%
  columns=fullflexible,
  keepspaces=true,
 commentstyle=\color{mygreen},
  escapeinside={(*@}{@*)}
}[keywords,comments,strings]%

\newcommand{\metavar}[1]{$#1$}
\newcommand{\metavars}[1]{$\MakeLowercase{#1}$}
\newcommand{\binop}{\diamond}
\newcommand{\transformsto}{\big\Downarrow}

\lstset{language=hojabr}

\lstMakeShortInline[columns=fixed, keepspaces=true, language=hojabr]!

\lstdefinestyle{sql_style}{
morekeywords={WITH,OVER},
commentstyle=\itshape\color{red}, keywordstyle=\codekwstyle{},
  aboveskip=1pt,
  belowskip=1pt,
  lineskip=-0.2pt,
, basicstyle=\footnotesize\ttfamily
}

\newenvironment{myboxed}
    {
    \begin{center}
    \begin{tabular}{|l|}
    \hline 
    }
    { 
    \\\hline
    \end{tabular} 
    \end{center}
    }

\newenvironment{hojabrcode}[0]
{ 
  \begin{lstlisting}
}
{ % End Code: Executed by \end{important}
  \end{lstlisting}
}

\begin{abstract}
Modern data analytics pipelines increasingly combine relational queries, graph processing, and tensor computation within a single application, but existing systems remain fragmented across paradigms, execution models, and research communities. This fragmentation results in repeated optimization efforts, limited interoperability, and strict separation between logical abstractions and physical execution strategies. 

We propose Hojabr as a unified declarative intermediate language to address this problem. Hojabr integrates relational algebra, tensor algebra, and constraint-based reasoning within a single higher-order algebraic framework, in which joins, aggregations, tensor contractions, and recursive computations are expressed uniformly. Physical choices, such as join algorithms, execution models, and sparse versus dense tensor representations, are handled as constraint-specialization decisions rather than as separate formalisms. Hojabr supports bidirectional translation with existing declarative languages, enabling programs to be both lowered into Hojabr for analysis and optimization and lifted back into their original declarative form. By making semantic, structural, and algebraic properties explicit, and by supporting extensibility across the compilation stack, Hojabr enables systematic reasoning and reuse of optimization techniques across database systems, machine learning frameworks, and compiler infrastructures.
\end{abstract}



\maketitle

\section{Introduction}
In recent years, there has been significant progress in artificial intelligence and data analytics, driven by advances in machine learning frameworks and large-scale data processing systems~\cite{DBLP:journals/pvldb/LowGKBGH12,DBLP:journals/pvldb/BoehmDEEMPRRSST16}. Improvements in hardware acceleration and co-processor–aware execution have further enabled increasingly complex analytical workloads~\cite{DBLP:conf/sigmod/BressFT16,jungmair2025towards}. Database systems have also evolved to support advanced analytics and tighter integration with machine learning pipelines, providing unified and optimized execution for data preparation and model training~\cite{DBLP:journals/pvldb/HellersteinRSWFGNWFLK12,DBLP:journals/pvldb/ZhaoAK24}. As a result, modern data-driven applications now operate at unprecedented scale and complexity.

This progress has made analytics pipelines increasingly heterogeneous. Modern workloads frequently combine relational queries with graph, tensor-centric analytics, and more expressive forms of reasoning within a single application~\cite{tian2019synergistic, gandhi2023tensor}. While individual systems are often highly optimized for specific classes of tasks, composing them into a coherent and efficient end-to-end pipeline remains challenging, particularly as systems strive to support multi-modal data and co-optimization across processing paradigms~\cite{gandhi2023tensor, he2022query}. This difficulty primarily stems from fragmentation across the paradigms and execution models that drive these workloads.

In addition to fragmentation across paradigms, there is substantial fragmentation within each paradigm. Modern database systems employ a wide range of execution models, including row- and column-oriented storage, vectorized execution, and compiled or code-generated query engines~\cite{abadi2013design, neumann2011efficiently, boncz2005monetdb,kersten2018everything}. Query processing further relies on a diverse set of join techniques, such as hash joins, merge joins, worst-case optimal joins~\cite{ngo2014skew}, free joins~\cite{wang2023free}, and specialized operators including group join~\cite{moerkotte2011accelerating} and diamond join~\cite{birler2024robust}. These techniques differ significantly in their internal representations, underlying assumptions, and optimization strategies~\cite{birler2024robust}, complicating unified query optimization and execution.

A similar form of fragmentation exists in tensor computation. Dense and sparse tensors are handled by largely separate systems, with different storage formats, kernels, and compiler infrastructures~\cite{chen2018tvm,kjolstad2017tensor,ragan2013halide}. Structured tensor algebra is typically optimized using techniques that differ from those used for unstructured sparse computation~\cite{spampinato2016basic,ghorbani2023compiling}. As a consequence, algebraically related computations are treated as distinct problems depending on the chosen representation, which limits the reuse of optimization techniques.

To make matters worse, progress in the database, machine learning, and compiler communities often occurs independently, and the transfer of techniques across communities is slow. Optimization methods developed in one setting (e.g., worst-case optimal joins or specialized tensor kernels), are rarely integrated into a shared framework. Instead, the knowledge base of optimization techniques remains scattered across systems and research communities.

\begin{figure}[t]
\includegraphics[width=\columnwidth]{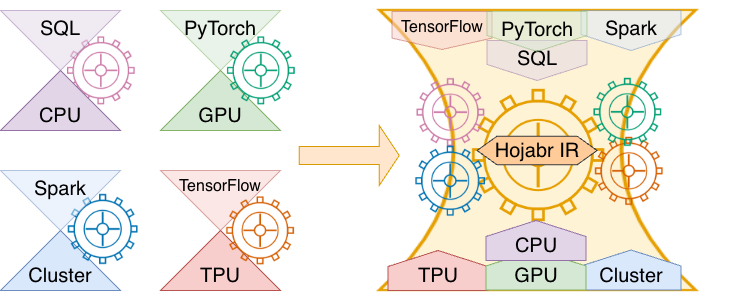}
\vspace{-0.8cm}
\caption{The architecture of state-of-the-art platforms versus Hojabr. Hojabr language enables interoperability between different systems and hardware architectures.}
\label{fig:arch}
\vspace{-0.4cm}
\end{figure}

We propose Hojabr (Higher-Order Joint Algebra for Bridging Representations) as a unified declarative intermediate language that addresses this fragmentation (Figure~\ref{fig:arch}). Hojabr combines relational algebra with techniques from tensor algebra and draws on mathematical frameworks such as linear programming to support constraint-based reasoning and optimization. In Hojabr, joins, aggregations, tensor contractions, and recursive computations are expressed within a single higher-order algebraic framework. Alternative execution strategies, such as different join algorithms or sparse versus dense tensor representations, are treated as constraint specialization choices rather than as separate formalisms.

Four principles guide the design of Hojabr.
First, higher-order relations enable the representation of relations, tensors, and their various physical data layouts as algebraic objects.
Second, constraint-aware semantics make semantic, structural, and algebraic properties explicit, including data types, cardinalities, degrees, functional dependencies, tensor shapes, and sparsity patterns.
Third, the language's extensibility allows the entire query compilation stack to be expressed, ranging from declarative specifications to low-level imperative code.
Finally, bidirectional bridging enables translation and reasoning across high-level DSLs and physical execution engines.

Hojabr not only provides a unified declarative intermediate language but also introduces new challenges and opportunities in the design of data analytics systems. By supporting bidirectional compilation, multi-paradigm reasoning, and execution across heterogeneous systems, Hojabr creates a foundation for reusable optimization, data science federation, and formally verified compilation. At the same time, this unified architecture raises new research problems in multi-paradigm optimization, cross-system interoperability, and large-scale verification.

The vision behind Hojabr is analogous to string theory in physics, which seeks a single mathematical framework capable of unifying fundamentally different physical forces. Similarly, Hojabr aims to unify diverse data analytics paradigms within one declarative algebraic theory. Different systems, representations, and execution strategies can be viewed as concrete realizations of a shared underlying structure, rather than as isolated and incompatible designs.


\section{Hojabr Language}
The Hojabr language is inspired by Datalog, Tensor Algebra, and Linear Programming. A Hojabr program consists of a collection of rules. Each rule has three components: (1) head, (2) expression, and (3) constraint.

\smartpara{Example} Consider the following program that applies a two-layer neural network to data produced by the join of two relations:

\begin{myboxed}
\begin{lstlisting}
J(i,a,b) := R(i, a), S(i, b)
X(i, j)  := v if J(i,a,b), match j case 0 -> v=a
                                   case 1 -> v=b
Z1(i, k) := x*w+b if x=X(i,j), w=W1(j,k), b=B1(k)
H1(i, k) := relu(v) if v=Z1(i, k)
Y(i)     := x*w+b if x=H1(i,j), w=W2(j), b=B2()
\end{lstlisting}
\end{myboxed}

The first rule computes the join of !R! and !S! and assigns the result to the relation !J!. The second rule converts the joined relation into a matrix with two columns: values associated with !a! are placed in column !0!, and values of !b! are placed in column !1!. The next rule computes !Z1! by performing a matrix multiplication between !X! and !W1! (which contains the weight matrix of the first layer), and then adds the bias term from !B1!. Then, !H1! is computed by applying the non-linear function !relu!. Finally, the predicted values are computed in !Y! by multiplying !H1! with !W2! and adding the bias term from !B2!.

\begin{figure}
\begin{tabular}{r r c l}
Rule & \metavar{R} & $::=$ & 
\metavar{A} \ruleaction{} \metavar{E} \codekw{if} \metavar{C}  $\mid$ 
\metavar{A} \ruleaction{} \metavar{C} 
\\
Constraint & \metavar{C} & $::=$ & 
\metavar{C}\code{,}  \metavar{C} $\mid$ 
\metavar{C}  \codekw{or}  \metavar{C} $\mid$ 
\codekw{not}\code{(}\metavar{C}\code{)} \\
& & $\mid$ & \metavar{R} $\mid$
\metavar{A} $\mid$ \metavar{E} $\theta$ \metavar{E} $\mid$ !CEI(!$\overline{C}$, $\overline{E}$!)!  \\
Expression & \metavar{E} & $::=$ & 
\metavar{A} $\mid$
\metavars{x} $\mid$ 
\metavar{E} $\binop$ \metavar{E} $\mid$ 
\metavar{\code{-}E} $\mid$ 
!EEI(!$\overline{E}$!)!  \\
Access & \metavar{A} & $::=$ & \metavar{X}\texttt{(}$\overline{E}$\texttt{)}...\texttt{(}$\overline{E}$\texttt{)} \\
Action & \ruleaction{} &  $::=$ & \code{:=}  $\mid$ \code{+=} $\mid$ \code{-=} $\mid$ \code{<-} \\
\end{tabular}
\vspace{-0.3cm}
\caption{The core grammar of Hojabr. $\theta$ and $\binop$ correspond to comparison and binary operations, respectively.}
\vspace{-0.3cm}
\end{figure}

\subsection{Higher-Order Relations}
\subsubsection{From Sets to Bags and Tensors} 

Datalog only supports relations with set-semantics~\cite{abiteboul1995foundations}, whereas Tensor Algebra supports real-valued tensors~\cite{vasilache2018tensor}. Inspired by work in database theory~\cite{abo2016faq,green2007provenance,abo2024convergence,koch2010incremental}, programming languages~\cite{DBLP:journals/pacmpl/ShaikhhaHSO22,shaikhha2019finally,dolan2013fun}, and high-performance computing~\cite{kepner2011graph,graph_blas}, we propose a semiring-based extension of relations.\footnote{Hojabr is not limited to the semiring structure; it can support more specific structures such as rings~\cite{koch2010incremental} or more general ones such as monoids~\cite{fegaras2000optimizing}.} Beyond standard set semantics, relations can include a payload that enables the specification of bags and tensors.

\smartpara{Example} When rule heads are only set-based relations, there is no need to include the expression part; only the constraints are required. In contrast, for rules whose head is a bag or a tensor, the expression part must be defined as well. In our running example, the first rule has a set-based relation !J!, whereas all other rules have a real-valued tensor head.

\subsubsection{Nesting} Another significant difference between Hojabr and Datalog is support for nested relations. This is enabled by allowing a relation to be applied to multiple argument lists. Although this representation is not required at the highest declarative level, it becomes essential for expressing join algorithms, as we will see later in Section~\ref{sec:usecases}.

\subsection{Constraint System}
Next, we present the constraint system, which, alongside expressions, forms the rules. Constraints can be either hard or soft; the former are used for verification (e.g., type checking), while the latter provides hints for optimization latter is used to provide hints for optimizations (e.g., cardinality information).
\subsubsection{Logical Atoms} 
The core of the constraint system is based on Datalog. Constraints can be conjunctions, disjunctions, and negations of atoms, as well as comparisons of expressions. Because relations are higher-order and can carry semiring payload values, both atoms and expressions can represent access to such relations.

\subsubsection{Nested Rules} To support nested rules, a rule itself can appear as a constraint. Such a constraint is satisfied if the head of the rule is non-empty. Section~\ref{sec:usecases} shows how nested rules can be used to model advanced join algorithms.

\subsubsection{Extension Interface}
Hojabr provides extension points for both expressions and constraints, denoted by !EEI! and !CEI!, respectively. Next, we describe different classes of !CEI!s.

\smartpara{Types} A central design decision in Hojabr is integrating its type system into the constraint system. This makes the type system both robust and flexible. The !type! constraint supports both attribute-level and relation-level typing (cf. Figure~\ref{fig:extension}). Moreover, types can be \textit{refined}~\cite{freeman1991refinement} by combining !type! with logical atoms, for example, by specifying a range constraint for an integer value. One can propagate these constraints by encoding type inference rules.

\smartpara{Structural Properties} Hojabr can encode database catalog information and tensor metadata. This includes cardinality, attribute degrees, tensor shapes, sparsity patterns, and attribute ordering. In addition to guiding optimization, these constraints can also encode the semantics needed for !ORDER BY! and !LIMIT! in databases: !ORDER BY! is modeled via !order!, while !LIMIT! can be modeled by using !card! to restrict the cardinality of the head of a rule. Furthermore, these constraints can encode schedule information used in frameworks such as Halide~\cite{ragan2013halide} and TACO~\cite{kjolstad2017tensor}.

\smartpara{Integrity Constraints} A final category corresponds to integrity constraints encoded in the database schema. These include functional dependencies (!fdep!), primary keys (!pkey!), and attribute uniqueness (!unique!).

\begin{figure}[t]
\begin{tabular}{|r l|}
\hline
\multicolumn{2}{|c|}{Constraint Extension Interface (\codemc{CEI})} \\ \hline
Types: & !type(!\metavars{x}!,!\metavar{T}!)! $\mid$ !type(!\metavar{X}!,!\metavar{T}!)!  \\
Structural: & 
!card(!\metavar{X}!,!$\overline{n}$!)! $\mid$ !deg(!\metavars{x}!,!\metavars{N}!)! $\mid$ !order(!$\overline{x}$!)! \\
Integrity: & 
!fdep(!$\overline{x}$!)(!\metavars{x}!)! $\mid$
!pkey(!$\overline{x}$!)! $\mid$ 
!unique(!$\overline{x}$!)! \\ \hline
\multicolumn{2}{|c|}{Expression Extension Interface (\codemc{EEI})} \\ \hline
Aggregates: & !avg(!\metavars{x}!)! $\mid$ !min(!\metavars{x}!)! $\mid$  !max(!\metavars{x}!)! $\mid$ !median(!\metavars{x}!)! \\ 
Math. Functions: & !sin(!\metavars{x}!)! $\mid$ !cos(!\metavars{x}!)! $\mid$ !relu(!\metavars{x}!)! $\mid$  !softmax(!\metavars{x}!)!  \\ \hline
\end{tabular}
\begin{tabular}{|r c l|}
\hline
\multicolumn{3}{|c|}{
\begin{minipage}{0.93\columnwidth}
\begin{center}
Syntactic Sugar
\end{center}
\end{minipage}
} \\ \hline
!n! $\theta_1$ !e! $\theta_2$ !m! &$\equiv$& !n! $\theta_1$ !e, e! $\theta_2$ !m! \\ \hline
!x in [t1, ..., tn]! &$\equiv$& !x=t1 or ... or x=tn! \\ \hline
\begin{lstlisting}
match t case t1 -> C1;
        ...
        case tn -> Cn;
\end{lstlisting} & 
$\equiv$ &
\begin{lstlisting}
(t=t1, C1) or 
   ...     or 
(t=tn, Cn)
\end{lstlisting} \\ \hline
!x: T! & $\equiv$ & !type(x, T)! \\ \hline
\end{tabular}
\vspace{-0.3cm}
\caption{A sample of extensions implemented for Hojabr.}
\label{fig:extension}
\vspace{-0.3cm}
\end{figure}

\subsection{Hojabr Sub Languages, a.k.a. Slangs}
We have seen the flexibility provided by Hojabr’s first-class citizens, including higher-order relations, !CEI!s, and !EEI!s. Inspired by the extensibility of MLIR’s intermediate languages (a.k.a. dialects)~\cite{lattner2021mlir}, Hojabr also provides sub-languages, which we refer to as \textit{slangs}. Hojabr’s extensibility features enable slangs to express queries from high-level specifications (e.g., at the level of SQL) down to a physical query plan and then to very low-level details (e.g., the implementation of individual and fused operators~\cite{shaikhha2018push,menon2017relaxed}).

\subsubsection{Declarative Slangs} Similar to declarative languages such as Datalog, SQL, and Tensor Algebra, Hojabr can express data analytics programs in a high-level manner. This makes Hojabr well suited to bidirectional translation into other declarative languages.

\subsubsection{Imperative Slangs} Hojabr can also specify lower-level imperative languages by allowing destructive rules. Rules support different actions (\ruleaction{}). Inspired by Temporel~\cite{DBLP:journals/pacmmod/ShaikhhaSSN24}, in addition to the declarative assignment (!:=!), Hojabr supports append (!+=!), remove (!-=!), and replace (!<-!).

\subsubsection{Iterative Slangs} Hojabr supports rule-level extensions to express more complex workloads. One example is recursion, which has received interest in both academia~\cite{shaikhha2026raqlet,10.1145/3299869.3324959,dedalus2011} and industry~\cite{aref2025rel,aref2015design,96b066e5eacd43b299251ec2a7e06a8b}. Many graph problems, linear algebra workloads, and database use cases require fix-point recursion that goes beyond the traditional query operators. Hojabr supports declarative fix-point recursion using rules that apply declarative assignment (!:=!) when a relation appears in both the head and the body. In addition, Hojabr supports an imperative specification of fixpoints by leveraging a looping construct~\cite{DBLP:journals/pacmmod/ShaikhhaSSN24}. Another example is support for a temporal dimension. This can be used to express models such as timely dataflow~\cite{murray2013naiad}, temporal databases~\cite{snodgrass1986temporal}, and iterative computations~\cite{DBLP:journals/pacmmod/ShaikhhaSSN24,murray2016incremental}.

\section{Theory and Systems of Everything}
\label{sec:usecases}
We have presented the core constructs of the Hojabr language. We have also shown how Hojabr slangs can cover a broader range of compilation stacks, compared to related work that relies on different languages for different concerns~\cite{DBLP:conf/sigmod/ShaikhhaKPBD016,jungmair2022designing,shaikhha2018building}. In this section, we provide additional detail on the expressive power of Hojabr across representative use cases in database systems and tensor systems.

\begin{figure}[t!]
\begin{tabular}{|l|l|}
\hline
$R\bowtie_{NLJ} S$  &
\begin{lstlisting}
Q(a,b,c) := R(a,b),S(b',c),(b=b')
\end{lstlisting} \\ \hline
$R\bowtie_{HJ} S$  &
\begin{lstlisting}
Rh(b)(a) := R(a,b)
Sh(b)(c) := S(b,c)
Q(a,b,c) := Rh(b)(a),Sh(b)(c)
\end{lstlisting} \\ \hline
$R\bowtie_{SMJ} S$  &
\begin{lstlisting}
Ro(b)(a) := R(a,b), order(b)
So(b)(c) := S(b,c), order(b)
Q(a,b,c) := Ro(b)(a), So(b)(c)
\end{lstlisting} \\ \hline
\end{tabular}
\vspace{-0.3cm}
\caption{The mapping of simple join operations in Hojabr.}
\label{fig:simplejoin}
\vspace{-0.3cm}
\end{figure}

\subsection{Database Systems}

\subsubsection{Hash and Sort-Merge Join} 
Hojabr can represent classic join operators natively (cf. Figure~\ref{fig:simplejoin}). Nested loop join is expressed as a single rule that joins two relations. A (Grace-like) hash join is expressed by constructing hash tables over the join keys of the two relations and then iterating over matching elements via those hash tables. Note that it is also valid to build a hash table on only one side (the build side) and to iterate directly over the elements of the other side (the probe side). Sort-merge join can be implemented by declaring the sorted column using the !order! constraint.

\subsubsection{Advanced Join Operators} 
The database literature has introduced several advanced join operators, with worst-case-optimal join algorithms as a primary line of work~\cite{wcojalg1,wcojalg2,graphflow,emptyheaded,leapfrog,ngo2014skew}. The key idea is to avoid materializing large intermediate relations by joining multiple relations simultaneously. Beyond their theoretical motivation, there is also sustained interest in integrating these operators into database systems by combining with binary join operators~\cite{graphflow,kaboli2025unified,wang2023free} and by exploiting parallelism~\cite{freitag2020adopting,wu2025honeycomb}.

Figure~\ref{fig:advjoin} illustrates how multiple state-of-the-art join algorithms can be implemented in Hojabr. The generic join algorithm~\cite{ngo2014skew} constructs tries (via higher-order relations) for each relation and then performs the join attribute-by-attribute. The free join algorithm~\cite{wang2023free} increases flexibility by relaxing the requirement to create a trie for every relation (e.g., relation !R! in the example does not require trie creation). Diamond join~\cite{birler2024robust} is another approach that uses lookup and expand operators to avoid unnecessary iterations. It relies on a hash table design with a dense collision list~\cite{birler2024simple}, which can be modeled in Hojabr using higher-order relations.

\begin{figure}[t!]
\begin{tabular}{|l|}
\hline 
\begin{lstlisting}
Q(x, a, b) := R(x, a), S(x, b), T(x)
\end{lstlisting} \\ \hline \hline
\begin{lstlisting}
// Generic join
Rh(x)(a) := R(x, a)
Sh(x)(b) := S(x, b)
Q(x,a,b) := Rh(x), (Rx:=Rh(x)), (Sx:=Sh(x)), 
  T(x), Rx(a), Sx(b)
\end{lstlisting} \\ \hline
\begin{lstlisting}
// Free join
Sh(x)(b) := S(x, b)
Q(x,a,b) := R(x, a), (Sx := Sh(x)), T(x), Sx(b)
\end{lstlisting}
\\ \hline 
\hline \hline
\begin{lstlisting}
Q(a,b,x1,x2,x3) := R1(a,b,x1), R2(a,x2), R3(b,x3)
\end{lstlisting} \\ \hline \hline
\begin{lstlisting}
// Diamond join
R2h(a)(x2) := R2(a, x2)
R3h(b)(x3) := R3(b, x3)
Q(a,b,x1,x2,x3) := R(a,b,x1), (R2a:=R2h(a)), 
  (R3b:=R3h(b)), R2a(x2), R3b(x3)
\end{lstlisting} \\ \hline
\end{tabular}
\caption{The mapping of advanced join operations in Hojabr.}
\label{fig:advjoin}
\end{figure}

\subsubsection{In-Database Analytics by UDF and Python Processing} 
In many workloads, queries require functionality beyond what SQL provides. Database systems address this gap via UDFs, which are often implemented using external languages such as Python~\cite{foufoulas2023efficient}. In practice, many data scientists bypass database systems and work directly with libraries such as Pandas DataFrames~\cite{pandas}, thereby losing opportunities for query optimization.

The Hojabr language is expressive enough to capture a wide range of functionality provided by UDFs and by such libraries. Moreover, beyond cost-based query optimization, compiler-based optimizations (e.g., inlining~\cite{shahrokhi2024pytond} and outlining~\cite{arch2024key}) can be expressed naturally on top of the language. This supports in-database analytics without incurring context-switching costs~\cite{fischer2022snakes}.

\subsubsection{Incremental Computation} In many settings, the data processing task operates over a stream of frequently changing data. In such cases, recomputing results from scratch can be unaffordable. Incremental processing techniques address this challenge by computing only the delta. These techniques are also used to compute fix-point programs efficiently in Datalog via semi-naive evaluation~\cite{abiteboul1995foundations}.

Hojabr’s algebraic nature makes it well-suited for applying existing incremental computation and incremental view maintenance (IVM) techniques~\cite{budiu2023dbsp,koch2014dbtoaster,murray2016incremental}. Moreover, thanks to its multi-paradigm design, Hojabr can potentially generalize prior efforts that apply IVM to linear algebra and hybrid workloads~\cite{nikolic2014linview,nikolic2018incremental,shaikhha2020synthesis}. Exploring this generalization is an important direction for future research.




\subsection{ML and Tensor Systems}
\subsubsection{Sparse Tensor Algebra}
Many applications require tensors with a large fraction of zero elements, including workloads in natural language processing and social network graphs. Such tensors can be modeled naturally using relations: a sparse tensor of order-$n$ is represented as a relation with $n+1$ columns, where the first $n$ columns encode the dimensions and the last column stores the value. This corresponds to the COO representation for sparse tensors.

\begin{figure}[t!]
\begin{tabular}{|l|}
\hline 
\multicolumn{1}{|c|}{Sparse tensor algebra} \\ \hline \hline
\begin{lstlisting}
// CSR representation (n,P,I,V)
B_CSR(i)(j) := V(p) if (0<=i<n),(p1=P(i)),
  (p2=P(i+1)), (p1<=p<p2), (j=I(p))
A(i)(j) := b*c if b=B_CSR(i)(j), c=C(j)
\end{lstlisting} \\ \hline \hline \hline
\multicolumn{1}{|c|}{Dense tensor algebra} \\ \hline \hline
\begin{lstlisting}
A(i)(j) := B(i)(j)*C(j) if card(B,n,m),card(C,m), 
  card(A,n,m), 0<=i<n, 0<=j<m
\end{lstlisting} \\ \hline \hline \hline
\multicolumn{1}{|c|}{Structured tensor algebra for 1-D convolution} \\ \hline \hline
\begin{lstlisting}
// Mapping of redundant and original elements 
B_R(i,j)(i',j') := card(B_O,n,_),(1<=j<=i), 
  (i<j+n), (j<n),(j'=0),(i'=i-j)
B(i)(j) := b if b=B_O(i)(j) or 
  B_R(i,j)(i',j'), b=B_O(i')(j')
A(i)(j) := b*c if b=B(i)(j) * c=C(j)
\end{lstlisting}
\\ \hline
\end{tabular}
\caption{Various types of tensor workloads in Hojabr. All of them correspond to matrix-vector multiplication.}
\label{fig:tensor}
\end{figure}

Prior work shows that state-of-the-art sparse tensor frameworks can outperform database systems by using more suitable storage layouts (e.g., CSR/CSC formats) and by employing algorithms closely related to WCOJ algorithms~\cite{shaikhha2024tensor,schleich2023optimizing,kovach2023indexed}. Hojabr can express these sparse data formats natively (Figure~\ref{fig:tensor}). Future work can investigate the potential in accommodating the recently proposed efficient database file formats~\cite{afroozeh2025fastlanes}.

\subsubsection{Dense Tensor Algebra}
Many tensor-algebra workloads involve dense tensors, including applications in computer vision and deep learning. Modeling dense tensors as database relations is typically impractical. Even for matrices (tensors of order-2), common approaches either rely on COO-style representations~\cite{blacher2023efficient} or inline all columns into relational schemas~\cite{shahrokhi2024pytond,blacher2022machine}. While Hojabr supports both approaches, neither is appropriate when the number of columns becomes large.

Hojabr’s data model provides a direct representation for dense computations (Figure~\ref{fig:tensor}). The tensor shape can be accessed via the !card! construct; in !card(B,n,m)!, !n! and !m! specify the number of rows and columns, respectively. This enables straightforward bidirectional translation between Hojabr and tensor algebra frameworks such as Halide~\cite{ragan2013halide}, TensorFlow~\cite{abadi2016tensorflow}, and PyTorch~\cite{paszke2019pytorch}.

\subsubsection{Structured Tensor Algebra}
Some operators are difficult to express efficiently using pure tensor algebra. Examples include convolution, which is widely used in machine learning, and the Kronecker product in quantum simulation workloads~\cite{tarabkhah2025synthesis}. These operators are commonly modeled via matrix computations over structured matrices, such as Toeplitz matrices~\cite{ghorbani2023compiling}.

Hojabr supports structured matrices and tensor operations over them, enabling convolution-like computations to be expressed natively in Hojabr (Figure~\ref{fig:tensor}). From this representation, one can translate to declarative structured tensor algebra languages such as STUR~\cite{ghorbani2023compiling} and LGen~\cite{spampinato2016basic}, or leverage the affine dialect of MLIR~\cite{ghorbani2025compressed} to generate specialized tensor kernels.

\subsubsection{Differentiable Programming} Computing the gradient of a function is a fundamental component of ML systems, since it is required to implement the optimization algorithms used during training. The programming languages and machine learning communities have developed extensive, systematic support for gradient computation, commonly referred to as differentiable programming~\cite{shaikhha2019efficient,jakob2022dr}. Because of its algebraic nature, Hojabr supports differentiable programming naturally. A key direction for future research is to investigate whether this support for differentiable programming can be extended to database operators.

\begin{figure}
\includegraphics[width=\columnwidth]{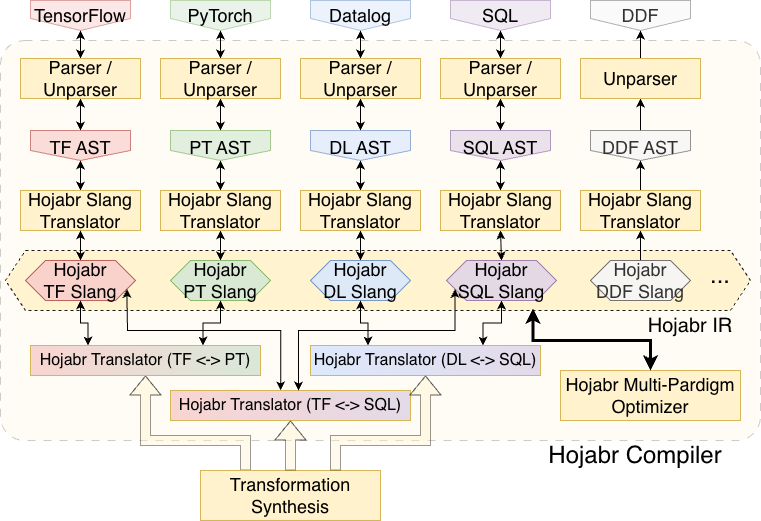}
\caption{The compilation workflow in Hojabr. Declarative languages have a bidirectional compilation pipeline, while lower-level languages can only be the target language. Hojabr programs can benefit from Hojabr's multi-paradigm optimizer or off-the-shelf optimizers of declarative languages.}
\label{fig:compiler}
\end{figure}

\section{Bidirectional Compilation}
In previous sections, we introduced the Hojabr language and its ability to express a wide range of database and AI workloads. In this section, we explain how recent advances in programming languages and compilers enable bidirectional translation between many existing declarative and imperative languages and Hojabr.

We decompose the bidirectional compilation problem into three components (cf. Figure~\ref{fig:compiler}). The first stage performs a bidirectional transformation between the source code and the AST of each language. We support this stage by providing a single interface that specifies both parsing and code generation (Section~\ref{sec:parserunparser}). The second stage performs a bidirectional transformation from the target AST to an appropriate Hojabr slang (Section~\ref{sec:toslang}). The third stage performs a bidirectional transformation across different Hojabr slangs (Section~\ref{sec:inslang}). In addition, Hojabr provides a one-directional translation to lower-level backends, such as differential dataflow (DDF), as shown in Figure~\ref{fig:compiler} (Section~\ref{sec:backend}).

\subsection{Parsing and Code Generation}
\label{sec:parserunparser}
The parser is the first component of a compiler and is responsible for transforming source code into a tree representation. Implementing parsers is often nontrivial and error-prone. Several frameworks provide specialized DSLs that allow developers to express parsers using a notation close to a language’s context-free grammar~\cite{cooper2022engineering}. Parser combinators are a successful realization of this idea in functional programming languages~\cite{hutton1992higher}.

Code generators (unparsers or pretty printers) can be viewed as the inverse of parsers: they translate a program’s tree representation back into a textual representation. Pattern matching in functional languages and string interpolation in modern languages provide practical support for writing code generators~\cite{hughes1995design}.

The structure of parser and code generator implementations is typically symmetric, since they are inverses of each other. Functional programming has leveraged this symmetry to provide libraries where both directions are specified once~\cite{rendel2010invertible}. This approach relies on bidirectional programming, a well-established programming languages technique~\cite{foster2009bidirectional}. Hojabr aims to benefit from bidirectional programming to implement parsers and code generators for multiple declarative languages while reducing maintenance costs.

\subsection{Transformation to/from Hojabr}
\label{sec:toslang}
Once the program is available as an AST, we translate it to the IR of an appropriate Hojabr slang. For each declarative language, we define a corresponding Hojabr slang that supports a straightforward one-to-one mapping. For example, dense tensor algebra (Figure~\ref{fig:tensor}) maps to a Hojabr slang that includes only the constructs required for dense tensor algebra; there is no need for attribute filtering, comparison, or negation.

Because this mapping is one-to-one, we can again leverage bidirectional programming~\cite{foster2007combinators}. With this approach, both translation directions are derived from a single specification, which eliminates the need to maintain separate implementations for forward and backward translation.

\subsection{Transformer Synthesis}
\label{sec:inslang}
After translating programs into Hojabr, we must support translation across Hojabr slangs. Bidirectional translation between slangs is more complex than in the previous two stages, because it may require global transformations that cannot be expressed as a single local transformation rule.

Our solution is to synthesize both translation directions from the specifications of the slangs. Each slang is specified as a set of logical constraints over the Hojabr syntax. For example, a slang for dense tensor algebra may allow only !card! constructs and relation access on the constraint side, and only multiplication and addition of values on the expression side.

Given these specifications and a general interpreter for Hojabr, the synthesis process can automatically derive transformations. Recent research shows that equality saturation~\cite{willsey2021egg,schneider2025slotted,tate2009equality} -- a technique that is conceptually similar to query optimization~\cite{schleich2023optimizing} -- can be leveraged to discover rewrite rules~\cite{nandi2021rewrite,pal2023equality}.

Synthesized transformations must be applied carefully. While the goal is to produce an equivalent program in the target Hojabr slang, many rewrites can substantially degrade performance. In practice, this motivates specialized transformations that account for performance constraints. An important research direction is to incorporate execution cost into the synthesis and selection of transformations (Section~\ref{sec:queryoptsas}).

\subsection{Backend Support}
\label{sec:backend}
We have shown how bidirectional compilation supports translation between different declarative languages and Hojabr slangs.

Hojabr is not limited to the backends exposed by existing systems. It can also generate code for lower-level IRs of existing systems (e.g., the sub-operator dialect of LingoDB~\cite{jungmair2023declarative} or the physical plan of DuckDB~\cite{raasveldt2019duckdb}) and for low-level data analytics libraries (e.g., differential data flow~\cite{mcsherry2013differential}). This allows Hojabr to reuse the optimization, parallelization, vectorization, and distributed processing capabilities of these frameworks.

Because the target languages are low-level, this translation is currently provided only in one direction. The reverse direction -- lifting low-level code into Hojabr slangs -- requires more advanced program synthesis techniques, which we discuss later in Section~\ref{sec:lifting}.

\section{Challenges and Opportunities}
\label{sec:challenges}
We have described the compilation workflow and how Hojabr supports bidirectional program transformation across languages. 
In this section, we discuss key challenges and opportunities enabled by the Hojabr architecture.

\subsection{Optimization as a Service}
\label{sec:queryoptsas}
Query optimization is one of the most (if not the most) complex components of a database system. In tensor processing systems, query optimization has only recently been leveraged~\cite {deeds2025galley}. Implementing optimizers across multiple systems typically requires substantial engineering effort.

Hojabr enables \emph{query optimization as a service}. Recent proposals such as Substrait~\cite{nadeau2022substrait} help make query optimization reusable across different systems~\cite{pedreira2023composable,alotaibi2024towards}. Hojabr achieves a similar goal through its language: different query plan languages can be translated to and from Hojabr, enabling the reuse of optimization across systems.

Instead of relying on an existing optimizer, one can also develop a stronger optimizer on top of Hojabr. This introduces two main challenges. \textbf{(1)~Multi-Paradigm Optimization:} since the input program may combine multiple paradigms, optimizing it is harder than optimizing a pure database query. \textbf{(2)~Multi-Paradigm Reasoning:} effective optimization requires reasoning across paradigms. For example, one needs to generalize cardinality estimation and integrity constraints to tensor and graph workloads.

\subsection{Data Science Federation}
\label{sec:federation}
Hojabr enables the execution of a single program on top of different systems. This is achieved by generalizing the idea of query federation; as opposed to query federation that only allows database queries to be executed on top of different systems, Hojabr allows data analytics programs consisting of snippets from different workloads to be executed on top existing systems. This is achieved by decomposing the input program into subprograms, routing each subprogram to the relevant system, and integrating the inputs/outputs of different systems.

Hojabr enables the execution of a single program across multiple systems. This generalizes the idea of query federation~\cite{sheth1990federated}. While classic query federation focuses on database queries across different database systems, Hojabr supports data analytics programs that include snippets from different workloads and executes them on top of existing systems. This is achieved by decomposing the input program into subprograms, routing each subprogram to a suitable system, and integrating the inputs and outputs across systems.

Related ideas have been explored in polystores, polyglot, and cross-platform processing systems~\cite{duggan2015bigdawg,dewitt2013split,kaoudi2022unified,kiehn2022polyglot,polystore_query_lang}. In addition, systems such as Weld~\cite{palkar2017weld}, IFAQ~\cite{shaikhha2020multi,shaikhha2021intermediate}, SDQL~\cite{DBLP:journals/pacmpl/ShaikhhaHSO22,shahrokhi2023building}, MatRel~\cite{matrel}, Raven~\cite{raven}, and LARA~\cite{laradb} aim to represent these workloads in a unified IR and execute them in a single runtime.

Generalized query federation introduces two primary challenges. \textbf{(1)~Boundary Identification:} to execute a program across multiple systems, the compiler must identify boundaries between subprograms. \textbf{(2)~Efficient Data Transfer:} interoperability requires transforming data into the formats expected by different systems. This can be addressed either by adopting common formats (e.g., Apache Arrow, as used in DataFusion~\cite{lamb2024apache}) or by generating efficient format conversions within Hojabr.

\subsection{Verified Optimization at the Large}
\label{sec:lifting}

A crucial advantage of Hojabr's mathematical foundation is its rigorous formal basis. Hojabr combines first-order logic from Datalog with arithmetic expressions in Tensor Algebra and linear programming constraints. In addition, structural and integrity constraints rely on core results from database theory~\cite{abiteboul1995foundations}.

This leads to the following opportunities and challenges. \textbf{(1)~Verified Compilation:} Hojabr can serve as a shared IR for formal specifications of query languages~\cite{DBLP:conf/cidr/0001L24} and tensor languages~\cite{kovach2023indexed,liu2024verified}. This can be enabled by implementing an interpreter for Hojabr in dependent-type-based proof assistants such as Coq~\cite{barras1997coq} or Lean~\cite{de2015lean}. This direction can build on the extensive literature on verifying query optimization and tensor compilation in these environments~\cite{liu2022verified,DBLP:journals/pacmpl/AuerbachHMSS17,DBLP:conf/pldi/ChuWCS17}. 
\textbf{(2)~Verified Lifting:} as discussed earlier, many existing codebases are written in low-level languages and are harder to translate into Hojabr than declarative programs. By combining program synthesis techniques with LLM-based approaches, it is possible to lift low-level codebases into Hojabr. This would enable the use of optimizations provided by database and tensor systems~\cite{cheung2013optimizing, li2025guided}.

\section{Conclusion}
This paper proposes Hojabr, a unified declarative intermediate language organized around multiple sub-languages, called slangs. Through bidirectional compilation, existing declarative languages are translated to and from appropriate slangs. This structure also frames the core challenges and opportunities: reusable optimization as a service across systems, principled cross-paradigm optimization between slangs, and large-scale verified compilation. 

Realizing this vision will require an implementation strategy that remains flexible across the stack and does not over-commit to a single technology base. We anticipate that components centered on bidirectional transformations over ASTs and Slangs will benefit from a functional implementation style, for instance, in Scala, where algebraic data types and pattern matching align well with compiler construction. Other parts, especially those related to hardware-relevant optimization pipelines, can naturally build on frameworks such as MLIR and its dialect and pass infrastructure.

\bibliographystyle{plainnat}
\bibliography{refs}
\end{document}